\documentclass[fleqn,10pt]{wlscirep}
\usepackage[utf8]{inputenc}
\usepackage[T1]{fontenc}
\newcommand{\ra}[1]{\renewcommand{\arraystretch}{#1}}

\title{Digital Traces of Brain Drain:\\Developers during the Russian Invasion of Ukraine}

\author[1,2,*]{Johannes Wachs}

\affil[1]{Vienna University of Economics and Business}
\affil[2]{Complexity Science Hub Vienna}

\affil[*]{johannes.wachs@wu.ac.at}

\begin{abstract} % abstract

The Russian invasion of Ukraine has caused large scale destruction, significant loss of life, and the displacement of millions of people. Besides those fleeing direct conflict in Ukraine, many individuals in Russia are also thought to have moved to third countries. In particular the exodus of skilled human capital, sometimes called brain drain, out of Russia may have a significant effect on the course of the war and the Russian economy in the long run. Yet quantifying brain drain, especially during crisis situations is generally difficult. This hinders our ability to understand its drivers and to anticipate its consequences. To address this gap, I draw on and extend a large scale dataset of the locations of highly active software developers collected in February 2021, one year before the invasion. Revisiting those developers that had been located in Russia in 2021, I confirm an ongoing exodus of developers from Russia in snapshots taken in June and November 2022. By November 11.1\% of Russian developers list a new country, compared with 2.8\% of developers from comparable countries in the region but not directly involved in the conflict. 13.2\% of Russian developers have obscured their location (vs. 2.4\% in the comparison set). Developers leaving Russia were significantly more active and central in the collaboration network than those who remain. This suggests that many of the most important developers have already left Russia. In some receiving countries the number of arrivals is significant: I estimate an increase in the number of local software developers of 42\% in Armenia, 60\% in Cyprus and 94\% in Georgia.
\end{abstract}

\begin{document}

\flushbottom
\maketitle

%Activating Crossborder Brokerage: ASQ Wang

\section*{Introduction}
The emigration of skilled individuals, sometimes called brain drain, is known to have important economic consequences for sending countries \cite{docquier2012globalization}. These effects can be both negative, for instance if the country cannot replace essential workers \cite{kremer1993ring}, or positive, for instance when leavers build social and economic networks between their origins and destinations \cite{saxenian2005brain}. Whether skilled emigration is on net good or bad for sending countries highly depends on the push and pull factors at play in specific cases. Push factors like war \cite{bang2013civil}, political instability \cite{ganguli2014scientific}, and terrorism \cite{dreher2011hit} cause shocks of emigration that are often irreversible. In these cases the sending country is less likely to benefit from positive externalities of its diaspora. 

At the same time it is often difficult, especially in volatile situations, to estimate how many people are leaving a country and what their destinations are. Lack of data on high-skilled emigration also hinders our understanding of differences between those who leave and those who stay. In recent years digital trace data has been used to provide insights into a variety of social, political, and economic phenomena \cite{conover2012partisan,smirnov2020estimating,glaeser2022}, including migration \cite{marquez2019segregation,gessler2021no,drouhot2022computational}. However, less is known about the dynamics of high skilled emigration and mobility during crises. 

One such crisis is the Russian invasion of Ukraine. While the flight of millions of Ukrainians is well-documented \cite{konstantinov2022impact}, less is known about emigration out of Russia and its supporting ally Belarus. Anecdotes and predictions of local professional organizations \cite{interfax22} suggest that many skilled individuals plan to or have already left these countries in response to the war \cite{aei22}, be it because of the effects of economic sanctions, fear of conscription, or moral opposition to the conflict. The Russian state, estimating that up to 10\% of IT specialists had left by December 2022 \cite{rumin2022russia}, has passed laws offering IT workers and firms tax breaks and benefits to entice them to stay \cite{braindrain2022russia}. 

A ``meaningful measurement'' \cite{lazer2021meaningful} of such migration would provide important insights into long-run effects of the war on the Russian and Belarusian economies that would be of interest for researchers and policymakers. First, analysts of the war itself can use robust estimates of Russian brain drain to better understand the resilience of the Russian economy, given that high skilled workers are an essential part of a modern economy. Second, potential receiving countries may consider adopting specific policies to attract and welcome talented \'emigr\'es. The rise in remote-friendly jobs since Covid-19 has already sparked a vigorous international competition for skilled workers \cite{sanchez2023home}. 

Finally, receiving countries can benefit from estimates of arrivals for the purposes of planning their effective integration into the labor force. While locals tend to have positive attitudes towards high skilled immigrants \cite{hainmueller2010attitudes}, such arrivals nevertheless often have trouble integrating optimally into local labor markets. They are more likely to be over-qualified for their positions \cite{quintini2011right}. When high skilled immigrants do integrate well, they can transfer valuable knowledge to local workers, significantly boosting local economies via spillovers \cite{kerr2010supply}.

Thus, this article seeks to provide novel insights into brain drain by studying the mobility open source software (OSS) developers residing in Russia and Belarus prior to the invasion of Ukraine in early 2022. Both countries have significant human capital in software and related technical fields \cite{wachs2022geography}. I first estimate the number of developers who have left those countries and compare these figures against a baseline generated from a comparison set of countries. Second, I observe heterogeneities in activity and collaborations between those who leave and those who stay. Finally, I report and interpret data on the destinations that developers move to.

To accomplish these aims, I use data from GitHub, the largest platform for online collaborative OSS development. I draw on and extend a large scale geographic census of highly active developers carried out in February 2021 \cite{wachs2022geography}, one year before the invasion. I revisit the profile pages of developers in the dataset, checking to see how many indicate a new geographic location. Specifically, I compare rates of migration of developers originally from Russia and Belarus with other countries in the region, including Ukraine. The primary empirical contribution of this work is to confirm that there has been a significant brain drain of software developers from Russia since the outbreak of hostilities. The data also indicates important heterogeneities between those who leave and those who stay behind: \'emigr\'es were previously significantly more active and more central in the collaboration network than their counterparts who stay behind. I also provide data on destination countries. More generally, this work demonstrates that digital trace data from professional platforms like GitHub can be used to provide relatively fast estimates of the magnitude and quality of brain drain in crises.

\section*{Background}
In this section I review related works on the effects of brain drain. I then provide justification why OSS developers are a good proxy for overall ICT activity in a country. Finally I provide information on the context of the analyses: the 2022 Russian invasion of Ukraine.

\subsection*{Related Works on Brain Drain}
As human capital is one of the primary factors driving economic growth \cite{mankiw1992contribution}, it is clear that the emigration of high-skilled workers can have a significant impact on an economy by decreasing its productivity. Second order effects are also thought to be significant. For example the productivity of remaining workers can suffer, especially if departing workers perform tasks that are highly complementary to other parts of the economy, or are difficult to substitute for \cite{kremer1993ring,docquier2012globalization,neffke2019value}. Wages of the remaining high-skilled workers may rise, increasing costs. The most pernicious effects, however, may be observed in the long run: human capital is more difficult to replace than physical capital. For example, academic emigration from Nazi Germany and deaths of academics in bombings during the Second World War diminished local research productivity for decades \cite{waldinger2016bombs}, while places which merely lost buildings and infrastructure recovered more quickly. 

Though there is evidence that brain drain can benefit sending countries through network effects (i.e. the transmission of information or capital back to the home country) \cite{saxenian2005brain} and increased incentives for local human capital formation \cite{docquier2012globalization}, it seems unlikely that these effects can manifest when talent is pushed out by socio-economic crisis or war. Indeed, previous work has shown that successful knowledge transfer between sending and receiving countries is highly conditional on people being embedded in both environments \cite{wang2015activating}. 

High-skilled immigrants also influence the economies of their destinations, bringing ideas and skills that can boost productivity and innovation \cite{moser2014german,ganguli2015immigration,miguelez2022migrant}. They can also shift the activity of locals \cite{ferrucci2020migration} and connect them with useful contacts back home \cite{saxenian2005brain,lHorincz2020global}. However, when a specific field is very competitive, high-skilled arrivals may push locals out of those labor markets \cite{borjas2015cognitive}. In the case of software development, an industry in which wages are high and jobs plentiful, the benefits of the former effects, sometimes called positive spillovers or externalities, likely outweigh the costs of the latter.

Less is known about heterogeneities between individuals who decide to leave and those who remain. Likely a complex mix of push and pull factors shape any individual's choice to emigrate. Any of a number of individual traits like personal motivation and ability or external factors like opportunities and costs can play a decisive role. For instance, it is known that social contacts abroad facilitate emigration: academics dismissed from their positions by the Nazis were significantly more likely to emigrate if they had collaborative ties to individuals who previously left \cite{becker2021persecution}. Such aspects also shape an \'emigr\'e's choice of destination \cite{docquier2012globalization}. However, while more skilled individuals may have greater opportunities abroad, they may have more reasons to stay: they may have built a strong local network and may be able to negotiate more favorable conditions. Indeed, as mentioned in the introduction, the Russian government is actively seeking to entice key individuals and firms to stay \cite{braindrain2022russia}.

\subsection*{Open Source Software activity as proxy for the ICT industry}
Here I argue that OSS plays a key role in the broader software industry and so serves as a useful proxy for the health and vigor of a country's ICT sector. OSS as a movement dates back to the 1970s and 80s \cite{eghbal2020working}. In the decades since, some of the most impactful and widely used software products are open source, including the Linux kernel and the Android operating system. A detailed discussion of why open source has been so successful is beyond the scope of this paper, but the open and transparent nature of OSS development, and its ability to integrate feedback and contributions from the crowd are key ingredients \cite{raymond1999cathedral}. By now OSS is often framed as a key infrastructure of our digital age \cite{eghbal2020working}.

This suggests that OSS matters in the global economy. For one, the software itself generates immense economic value as public goods that anyone can use \cite{greenstein2014digital}. But there are significant second order effects that accrue to individuals, firms, and places that are active in OSS. Individuals contribute to OSS for many reasons, including for their own enjoyment, but tend to accrue real economic benefits from doing so \cite{lerner2002some}. For example, a prolific GitHub page gives a developer a significant value in the software labor market \cite{marlow2013activity,papoutsoglou2019extracting}; some even earn a living from crowdfunded sponsorships of their work \cite{overney2020not}. Firms become more productive from contributing to OSS \cite{nagle2018learning} and by using it \cite{nagle2019open}. OSS contributions are also valuable signals of information for investors, allowing them to observe and verify the quality of software written by individuals and firms they are considering investing in \cite{wright2021open}. These factors and others suggest why countries benefit at the macro level from local OSS contributions \cite{blind2021impact}. OSS activity is strongly correlated with productivity, innovation, economic complexity, and growth outcomes at the national and regional levels \cite{wachs2022geography}. In other words, even though OSS activity is perhaps uniquely amenable to online and remote collaboration \cite{goldbeck2022bit}, it still matters where OSS is made. In these ways observed OSS activity is a strong proxy for knowledge-intensive and productive ICT activity.  

\subsection*{Context}
Russia has occupied Crimea and supported insurgencies in two eastern Ukrainian regions since 2014. Russia's invasion on February 24th, 2022 nevertheless represented a significant escalation of the scope and scale of the conflict. Despite observed troop movements and reports from western intelligence services that an attack was imminent, the general public throughout Europe was rather surprised by the invasion \cite{berlinschi2022rallying}. Part of the invasion went through Belarus, which provided logistical support. Anecdotes of a sharp rise in Russian emigration, especially among high-skilled workers, quickly emerged \cite{aei22}. Besides opposition to the war, fear of eventual conscription was likely a significant motive. Although Russia initiated partial mobilization only several months after the invasion, the invasion itself increased the risk of such an event in the near future; as young men are significantly overrepresented in the OSS community (often estimated at around 90\% of contributors \cite{terrell2017gender}), this would impact the majority of developers. The invasion also likely cemented existing perceptions among skilled 
young people in Russia that prospects for political liberalization are growing dimmer \cite{demintseva2021understanding}. External forces also may have played a role in migration decisions. For example countries may restrict immigration from Russia in response to the war. Indeed, as of August 2022, the EU has already considered suspending its arrangements with Russia for simplified visa procedures.

\begin{figure}[t]
    \includegraphics[width=0.8\textwidth]{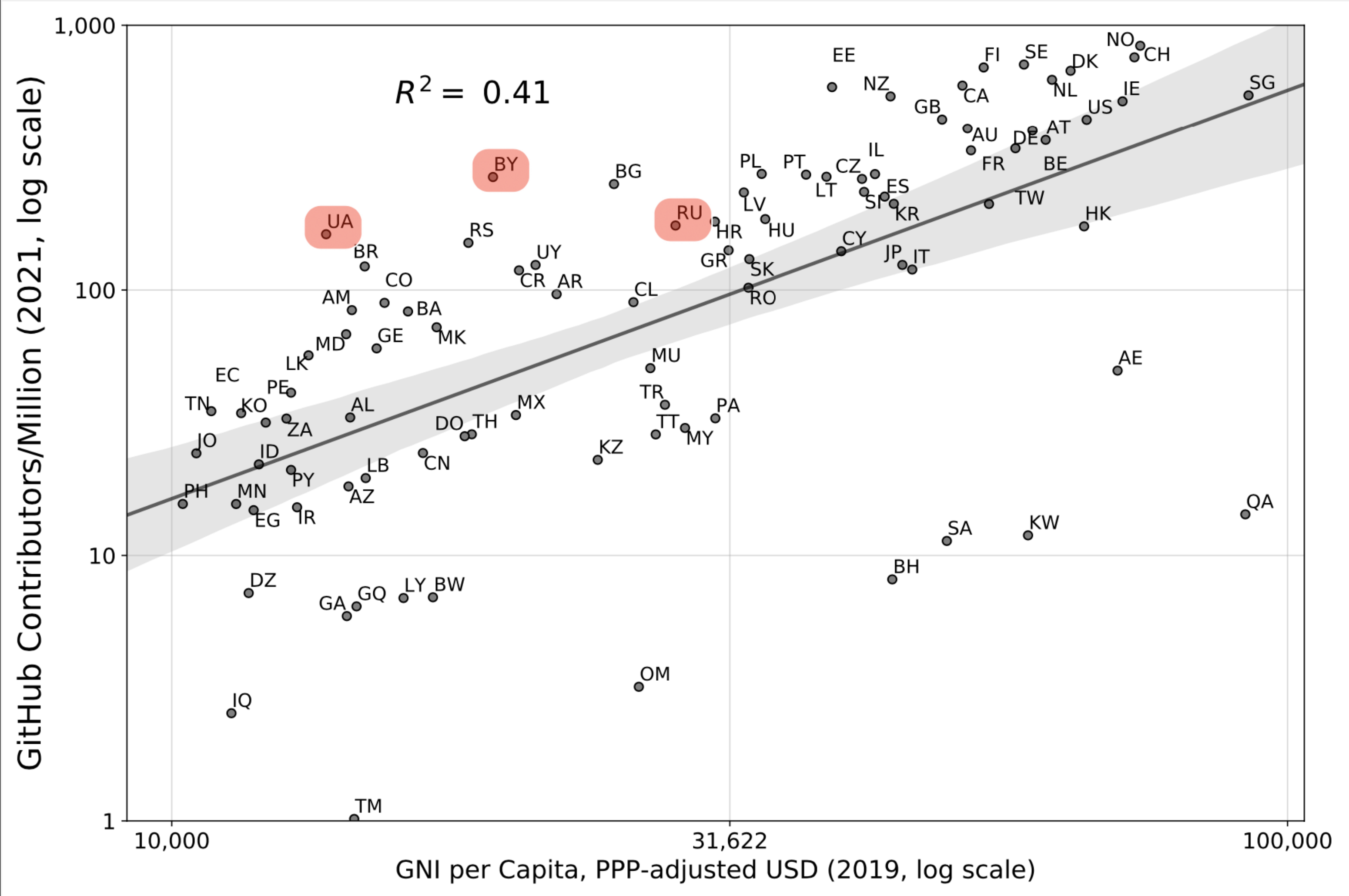}
    \caption{The relationship between national economic development and highly active developers on GitHub per capita on a double logarithmic scale, adapted from \cite{wachs2022geography}. Ukraine (UA), Belarus (BY), and Russia (RU), highlighted in red, are significant positive outliers from the trend line, meaning that they have many more OSS developers than expected given their levels of economic development.}
    \label{fig:dev_vs_gh}
\end{figure}

High-skilled emigration of software developers is likely to cause the Russian economy significant trouble. Russia and other post Soviet states have advanced information technology sectors \cite{biagioli2019russia}, owing to a strong tradition of technical and engineering education. For instance, 7.1\% of gross Russian exports are categorized as information and communication technologies by the Atlas of Economic Complexity \cite{hausmann2014atlas}. Indeed, previous work comparing economic development and per capita contributions to Open Source Software found that Russia, Ukraine, and Belarus were significant positive outliers \cite{wachs2022geography}, see Figure~\ref{fig:dev_vs_gh}. People with skills in software are a key input to growing sectors of the digital economy \cite{sturgeon2021upgrading}, a fact underlined by persistently high wages \cite{breaux20212021} in the industry, and the high share of US H1B visas going to software engineers.

Belarus's role in the invasion also merits discussion. Russian troops used Belarus as a staging ground for part of its invasion, notably its attacks on Kyiv, the Ukrainian capital. Weapons on Belarusian territory have fired on Ukraine, and so the country is typically thought to be an accomplice if not co-belligerent. The Belarusian presidential elections in 2020 and the subsequent protests and large scale demonstrations continuing into 2021 likely impact the analyses of Belarusian developers, as these events are thought to have lead to significant brain drain themselves \cite{klysinski2021protest}.

\section*{Data and Methods}
In this section I outline the original dataset surveying the global geographic distribution of active GitHub developers in February 2021. I then describe how I revisited the profiles of developers in June 2022, in the aftermath of the Russian invasion, and in November 2022, following Russia's partial mobilization order, to generate data on changes in developer locations. Finally I give a brief overview of the methods used in the analyses.

\subsection*{The Geography of OSS in 2021}
A recent work mapped the geographic distribution of OSS developers using data from GitHub \cite{wachs2022geography}. That paper used data from GHArchive, a database of activity on GitHub, to identify individuals highly active on the platform. In particular, the dataset consists of around 1.1 million developers who made at least 100 commits to public GitHub projects across the 2019 and 2020. Commits are elemental code contributions to OSS projects. These developers were then geolocated using three data traces observed in February 2021: the free text location field in their profile pages, their use of geographically identifying email suffixes, and the location associated with their linked Twitter accounts, if they have one. Using the Bing Maps API the authors could geolocate around half of the active developers to at least the country level. The resulting dataset gives a comprehensive overview of where OSS developers live at one moment in early 2021. It also includes information on the number of contributions made by each developer and to which projects they were made. I will use this extra information to compare the activity and collaboration network position of developers who remain in Russia with those who have left since the invasion.

\subsection*{Revisiting Eastern European Developers in 2022}
Focusing on Russia, Belarus, Ukraine, and comparable countries in the region\footnote{Estonia, Latvia, Lithuania, Poland, Czechia, Slovakia, Hungary, Romania, Bulgaria, and Serbia.}, I revisited the profile pages of around 45,000 developers in late June 2022. I restricted my attention to those developers geolocated by their metadata on GitHub in the previous work, ignoring Twitter data and email commit suffixes. Specifically, I used the GitHub REST API to query up-to-date user data for each developer, including their plain-text location field.

There were several possible outcomes for each individual. First, an individual may have deleted their account since February 2021. Second, an individual may have the same location string in their profile; in this case the individual was simply assigned to the same geolocation as before. Finally, an individual may have updated or deleted their location string. When this new string was identical to a location already processed before, for instance ``Moscow, Russia'', the individual was assigned to that geolocation. If the string was new, I used the Microsoft Bing API, as was done in 2021, to geolocate it. The Bing API handles multiple languages and returns multiple suggested locations, ranked by a likelihood calculated by Bing. Following the procedure carried out in 2021, I assigned these users to the location with the highest likelihood as determined by the API. 

The resulting dataset includes around 45,000 observations. Each observation is a developer geolocated in one of the countries listed above in early 2021. I record their original plain text location and Bing Maps geocode (country and, when available region or city), and the updated equivalent on June 22nd, 2022, and again on November 8th, 2022. I note account deletions and cases in which an individual removes their location information. I also record the number of commits they made in 2019-2020, and a list of projects they made these commits to. Given the sensitive nature of this data, an anonymized version of the dataset (removing developer login and plain text location, keeping only the geocoded data and the number of contributions made in 2019-2020) is available here: \url{https://github.com/johanneswachs/ru_braindrain_data}.

\subsection*{Methods}
While the first part of the results simply present count statistics on developer movement, I employ various methods to study heterogeneities between leavers and remainers. First I study the differences in prior activity, counting commits. I compare the distribution of previous commits by leavers and remainers visually, then report means and medians, testing for significance of the former using a Mann-Whitney U test. 

I use network science methods to study differences in collaboration methods between leavers and remainers. Specifically I construct a collaboration network of developers by connecting developers if they contribute to the same repo in the two years of 2019 and 2020 \cite{lima2014coding}. I compare differences in leaver and remainer degrees (that is, the number of connections), and their representation in the largest connected component of this network. To quantify the extent to which leavers are more or less central in the Russian collaboration network, I measure the change in the network's overall connectivity if they are removed, compared with removing a random subset of developers of the same cardinality. I also report differences in the number of connections with developers in other countries between leavers and remainers. Finally, I present summary statistics on destinations and show that leavers are more likely to have previous ties to developers in their destination country. 

\section*{Results and Analyses}

\subsection*{International Comparison}
I first compare the geographic mobility of developers originally located in various Eastern European countries. Comparing June 2022 and then November 2022 against February 2021, I report how many developers have delete their profiles, how many provide an invalid (i.e. ``the moon'') or no location, and how many signal a location in a new country. The results are described in Table 1.

\begin{table}
\ra{1.2}  
\begin{tabular}{llcccccccc}
  &  & \multicolumn{2}{c}{Profile Deleted (A)} & \multicolumn{2}{c}{Invalid/No Loc. (B)} & \multicolumn{2}{c}{New Country (C)} & \multicolumn{2}{c}{A+B+C}  \\
\cmidrule(lr){3-10}
Country & Devs Feb21  & Jun22 & Nov22  & Jun22 & Nov22  & Jun22 & Nov22 &Jun22 & Nov22   \\
\cmidrule(lr){1-1}\cmidrule(lr){2-2}\cmidrule(lr){3-4}\cmidrule(lr){5-6}\cmidrule(lr){7-8}\cmidrule(lr){9-10}
   Russia &   15,543 &  3.0\% & 3.5\% &  11.3\% & 13.2\% &    7.4\% & 11.1\% &    21.7\% & \textbf{27.8\%} \\
  Belarus &   2,343 &   3.4\% & 4.1\% & 7.0\% & 8.8\% &  12.2\%&  16.8\% &  22.6\%&  29.7\% \\
  Ukraine &   6,939 &   3.2\% & 3.9\% & 2.0\% & 2.7\% &  3.0\%& 4.2\% &   8.1\%&  10.8\% \\
  \midrule
 Estonia &    600 &   1.7\% & 1.8\% &   2.6\% &  2.9\% &  5.3\%&  6.2\% &  9.6\% & 10.9\% \\
 Latvia &    371 &  2.7\% &   2.7\% &   1.7\% &  3.3\% &  2.7\% & 2.7\% &   7.1\% & 8.7\% \\
Lithuania &    683 &  2.5\% &  2.8\% & 2.1\% & 2.4\% & 1.8\%&   2.2\% & 6.4\%&  7.4\% \\
   Poland &    8,865 &  2.4\% &  2.9\% & 1.8\% & 2.2\% &   1.6\% &   2.1\% & 5.8\%  & 7.2\% \\
  Czechia &    2,771 &  1.7\% &  2.1\% & 1.7\% &  2.2\% &   2.8\%&   3.5\% &  6.2\%&  7.8\% \\
 Slovakia &    620 &  2.4\%&   2.6\% &  2.7\%&  3.0\% &   3.9\%& 4.8\% & 8.9\% &  10.4\% \\
  Hungary &    1,616 &  1.1\%& 1.4\% &  2.1\% & 2.6\% & 3.8\%& 4.5\% & 7.0\% & 8.6\% \\
  Romania &    1,820 &    1.8\% & 2.1\% & 2.3\%& 2.5\% &  2.9\% & 3.5\% &  6.9\%& 8.2\% \\
 Bulgaria &    1,509 &    2.3\%& 2.7\% & 1.4\% &  1.8\% &  1.3\%& 1.9\% & 4.9\% &  6.3\% \\
 Serbia &     953 &  2.7\%&  3.1\% &  2.2\% &  3.2\% & 2.6\%& 3.5\% & 7.5\%&  9.8\% \\
   \midrule
 ex-R/B/U &  19,808 &  2.1\%& 2.5\% & 1.9\% & 2.4\% & 2.3\% & 2.8\% &  6.3\%& 7.8\% \\
\bottomrule
\end{tabular}
\label{tab:summary_nov}
\caption{Statistics on OSS developers in select CEE countries, originally observed in February 2021. Developer profiles on GitHub are revisited in June 2022, following the Russian invasion of Ukraine, and in November 2022, following Russia's partial mobilization. Besides the case when a developer lists the same country, profiles are either deleted, list an invalid (i.e. ``the moon'') or no location, or list a location in a new country. The final column sums the previous three. The final row considers the aggregated statistics of all countries besides Russia, Belarus, and Ukraine.}
\end{table}

By June over one in five developers previously located in Russian (21.7\%) and Belarus (22.6\%) could no longer be geolocated there. This is three to four times higher than the baseline rate (6.3\%). I find a higher ratio by November, following Russia's partial mobilization order: 27.8\% of developers in Russia and 29.7\% of developers in Belarus in 2021 could no longer be located there, compared with 7.8\% of developers in comparison countries.

Indeed, by November 2022 Russia-based developers were about four times as likely to list a new country than the baseline (11.1\% vs 2.8\%). They are also more likely to have deleted their profiles (3.5\% vs 2.5\%) or to have obscured their location (13.2\% vs 2.4\%). This last observation is especially striking: over one in eight previously geolocatable Russian developers has obscured their location. One explanation is that signaling that one lives in Russia after the invasion has a social or economic cost. Another is that they have left Russia and have not yet settled on a final destination or have done so and do not wish to signal it. Unfortunately I cannot tell which of these developers remain in Russia and which have left. 

These summary statistics suggest that there has been a significant emigration of Russian (and Belarusian) software developers between early 2021 and November 2022. Interestingly, Ukrainian developers do not appear to have emigrated in much greater numbers than the regional benchmark. This may be explained by the ban on young men from leaving the country or a widespread desire to participate in its defense.

\subsection*{Heterogeneities between leavers and remainers}
I now compare two populations of Russian developers which I call leavers and remainers. Leavers are those who list a new country in their GitHub profiles. Remainers are those that still signal that they live in Russia. I examine differences between the two in activity, in their regional geographic location, and in position in the pre-invasion Russian developer collaboration network.

\begin{figure}[t]
    \includegraphics[width=0.7\textwidth]{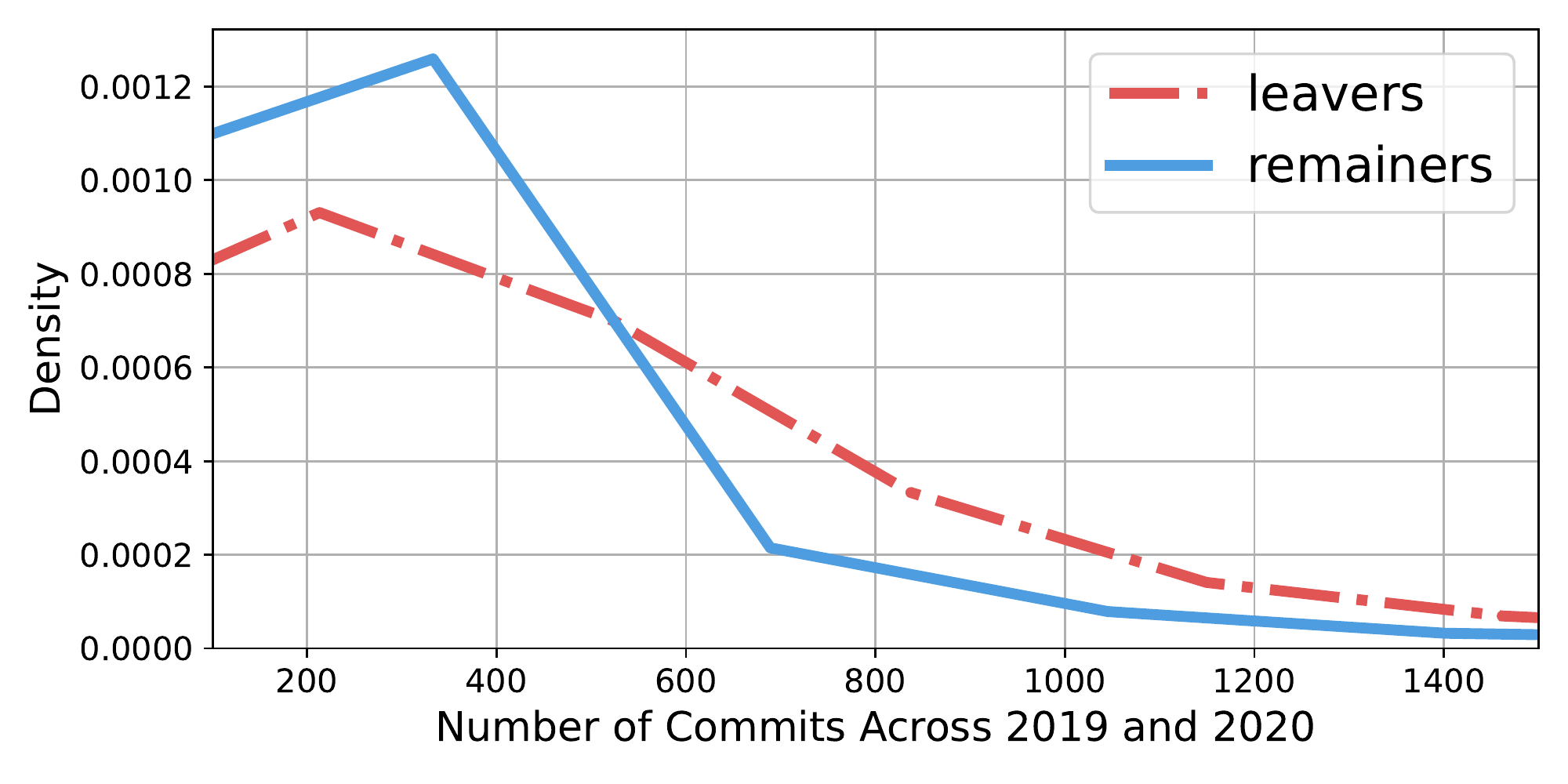}
    \caption{Normalized distributions of activity 2019-2020 of highly active developers geolocated in Russia in early 2021. The red dotted line is the distribution of developers who list a non-Russian location in November 2022. The blue solid line is the same for developers who still list a Russian location. Note that the threshold for inclusion in the census was making at least 100 commits in 2019 and 2020, hence the distributions are truncated on the left at 100.}
    \label{fig:activity_diffs}
\end{figure}

\subsubsection*{Activity}
I can quantify the activity level of different developers by counting the number of commits they made in 2019-2020. Commits are elemental contributions of code to OSS libraries \cite{dabbish2012social}. Recall that the threshold for inclusion in this dataset was making 100 commits across these two years. I find that leavers were significantly more active, see Figure~\ref{fig:activity_diffs} for a comparison of the normalized distributions of activity between the two groups. Leavers averaged 418 commits (median: 204), while remainers averaged 301 (median: 171). This analysis also provides a way to measure the cumulative impact of the departure of developers on the Russian IT landscape. Confirmed \'emigr\'es account for 11.1\% of developers but make 14.0\% of commits in 2019-2020. These findings are qualitatively robust to filtering for developers who made at least 200 or 500 commits in the previous period, as reported in the supplementary appendix.

\subsubsection*{Geography}
Developers originally from St. Petersburg and Moscow, the two leading cities of Russia, likely have larger, more diverse networks than their counterparts from more peripheral regions \cite{eagle2010network}. Such connections are thought to be invaluable for individuals leaving a country in crisis \cite{becker2021persecution}. They may also have more financial resources they can use to move. On the other hand, developers working in the main hubs of Russia may be more established and have more to lose by leaving. Among Russia-based developers for which a subnational-geolocation is available in 2021, I find a small, statistically significant difference: 15.2\% of developers originally from these two cities now list another country compared with 11.7\% of those from other parts of Russia (MW-U $13,778,529$; $p<.001$).

\subsubsection*{Collaboration Networks}
OSS development, indeed software development in general, is a highly collaborative endeavor \cite{dabbish2012social,celinska2018coding,zoller2020topology}. To study differences in the collaboration patterns of the leavers and remainers, I construct the collaboration network among Russian developers by first creating the bipartite graph of developers and the GitHub repositories they make at least two commits to in the years 2019-2020. Following previous work \cite{lima2014coding,celinska2018coding} I then project this network onto the developers, connecting them with an edge if they contribute to the same project. 

\begin{figure}[t]
    \includegraphics[width=0.8\textwidth]{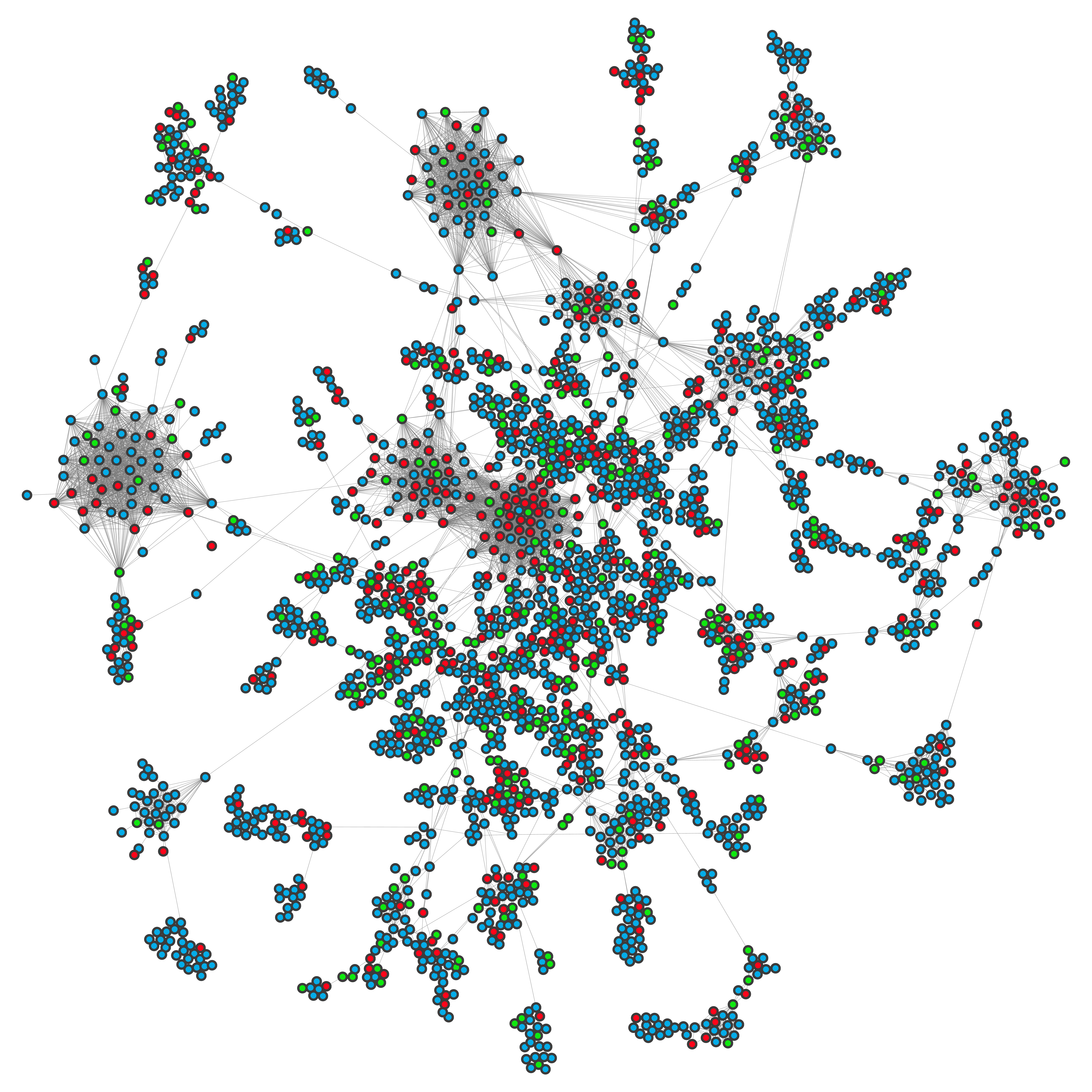}
    \caption{The largest connected component of the collaboration network of OSS developers located in Russia in early 2021. Two developers are connected by an edge if they contribute to the same repository. Nodes colored red are geolocated to other countries besides Russia in November 2022. Blue nodes are still in Russia, and green nodes have obscured their locations.}
    \label{fig:gcc}
\end{figure}

The resulting network gives a simple overview of the collaborations within the Russian OSS ecosystem. It consists of 4935 non-isolated nodes and 12988 edges and is rather fragmented: only 2208 nodes are in the largest connected component. I visualize the largest connected component in Figure~\ref{fig:gcc}, coloring nodes based on their updated location. Blue nodes can still be located in Russia, red nodes are geolocated in another country, and green nodes have obscured their location.

\begin{figure}[t]
    \includegraphics[width=0.7\textwidth]{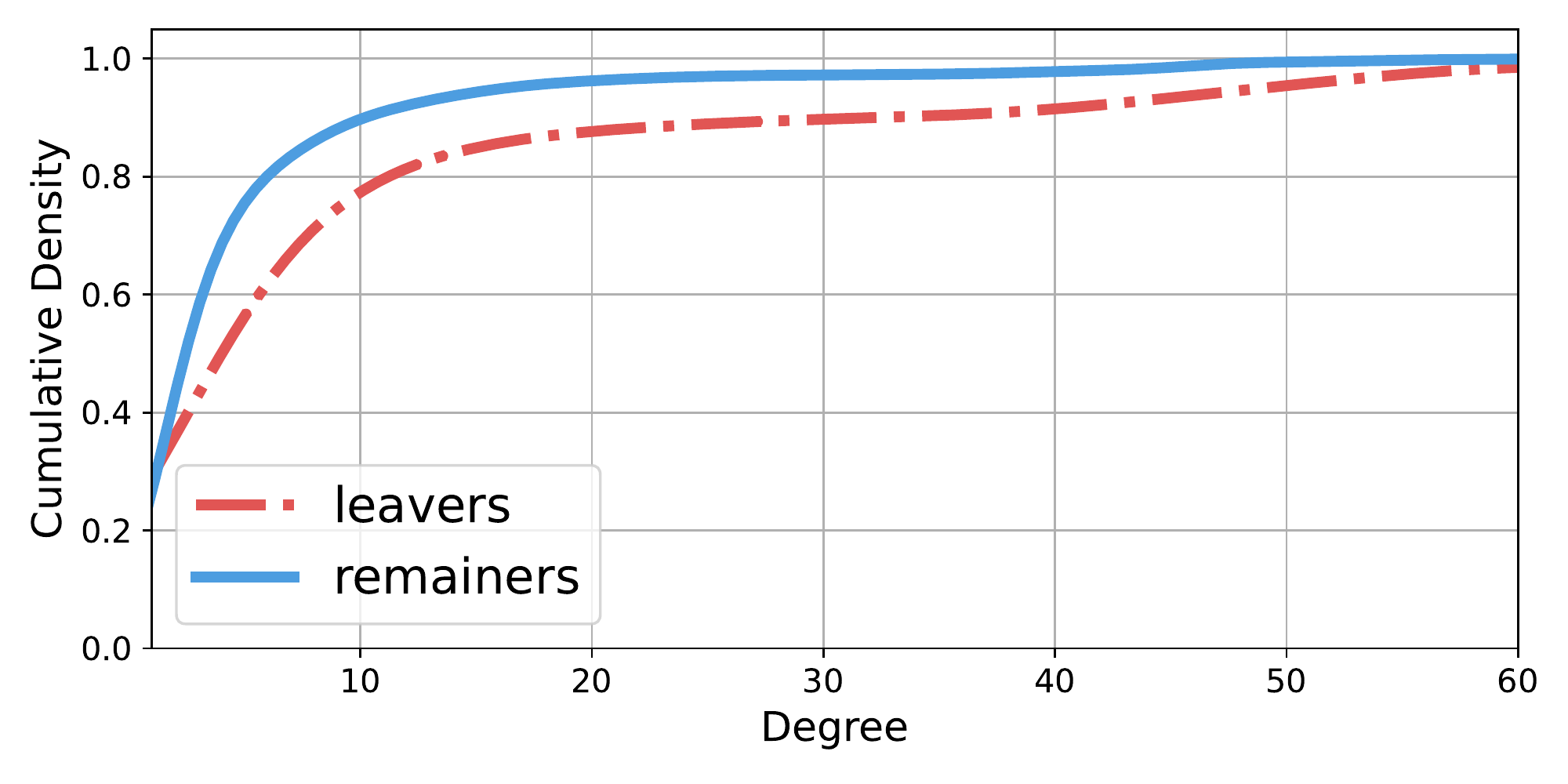}
    \caption{The cumulative distributions of the number collaborations of leavers and remainers in the Russian community on GitHub, mapped using activity in 2019 and 2020 and including only developers with at least one connection.}
    \label{fig:deg_dist}
\end{figure}

The visualization hints that leavers may be more connected and more central than their remainer counterparts. Indeed the statistics bear this out. In the full network, leavers have on average 3.6 collaborators (standard deviation 10.7), while remainers average 1.4 (s.d. 4.9). Both distributions are rather skew, so I plot the cumulative distribution of the number of connections of leavers and remainers, respectively, with at least one connection in Figure \ref{fig:deg_dist}. The distributions suggest that the difference in means is not driven by a small number of outliers. For instance: roughly 20\% of leavers have at least 10 collaborators, while only about 10\% of remainers do.

The network visualization also suggests that leavers play an important structural role in the network. That is to say, leavers are not only connected with more collaborators, but they are also occupying more important positions in the network. I quantify the structural importance of leaving developers is via a network robustness exercise, applying a method originally used to study the resilience of protein interaction networks \cite{zitnik2019evolution}. Given a network $G$ with $N$ nodes partitioned into a set of connected components $\{C_{1},C_{2},...C_{k}\}$, the normalized Shannon entropy of $G$ can be defined as:

$$ H(G) = - \dfrac{1}{\log(N)} \sum_{i=1}^{k} p_{i}\log p_{i},$$

where $p_{i} = C_{i}/N$ is the share of nodes in the $i$-th connected component. In a fragmented network, in which nodes are split into small, similarly sized connected components, this entropy score tends to 1. In a network in which more nodes are in a single connected component, the score tends to 0. 

\begin{figure}[t]
    \includegraphics[width=0.8\textwidth]{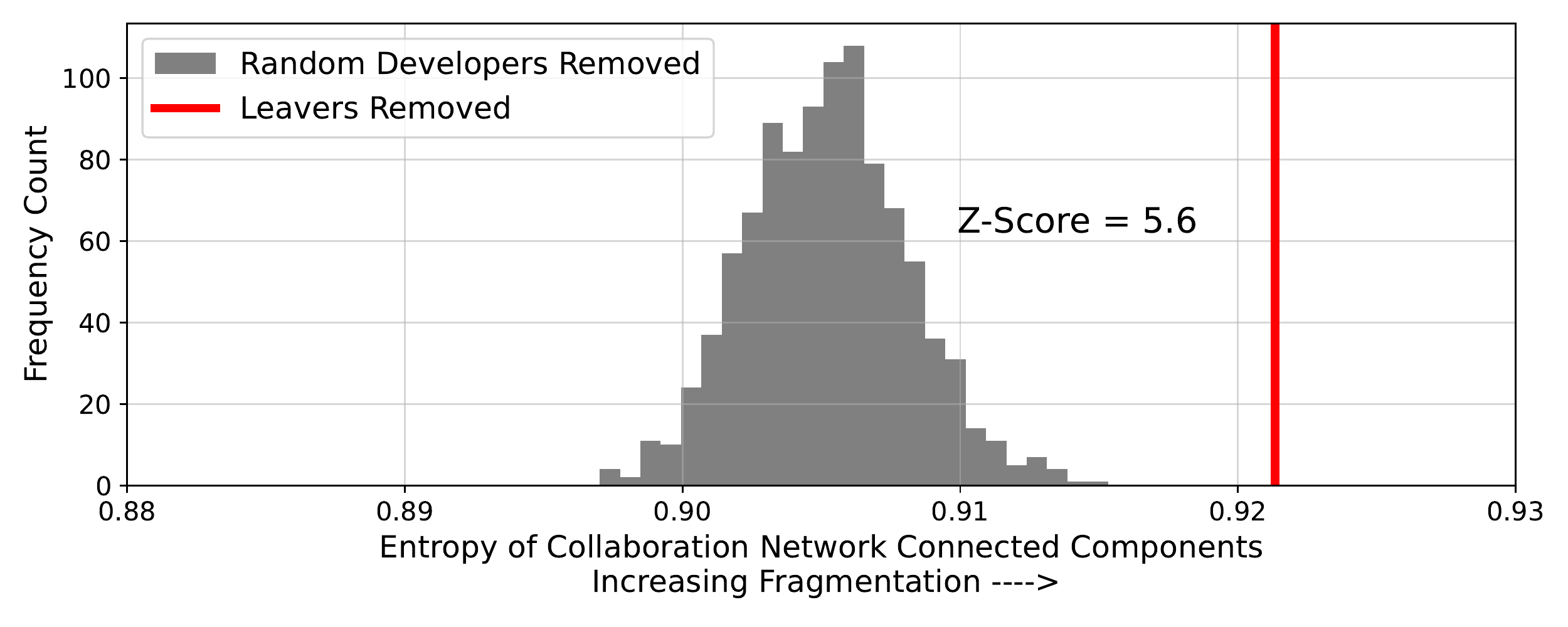}
    \caption{The normalized Shannon entropy of the Russian software developer collaboration network when removing leavers (in red) and a random subset of developers of the same size (1000 times, distribution in grey). Higher scores indicate greater levels of fragmentation. The Z-score compares the targeted removal with random removal, and indicates that leavers play a distinguished role in overall network connectivity.}
    \label{fig:ent}
\end{figure}

I calculate this score for two versions of the developer collaboration network. First, I remove all of the leavers from the network. Second, I remove a random subset of developers equal in size to the number of leavers. The second calculation serves as a null model which benchmarks the effect of developer removal on network fragmentation. I repeat the second calculation 1,000 times and obtain a distribution of entropies under the null model, which I compare to the entropy observed when removing leavers using a Z-score. The results are summarized in Figure \ref{fig:ent}.

The Z-score of 5.6 indicates that leavers play a distinguished role in the collaboration network. Their removal leads to a significantly more fragmented network than under the null model of removing the same number of developers, randomly selected. This suggests that they are playing structurally important roles in the Russian OSS community. 

As many developers may contribute a given project, it may be the case that two developers contributing to a project may not really be collaborating with each other. I therefore rerun the same analyses on a collaboration network constructed using only projects with 10 or fewer contributors. I find similar results, reported in the supplementary appendix.

\subsection*{Network position and emigration}
Having established that leavers are better connected and tend to occupy more central positions in the Russia-based developer collaboration network before the war, I now ask whether network connections outside of Russia are associated with the likelihood a developer leaves. To do so, I repeat the construction of the collaboration network, this time including all accounts geolocated in 2021 which contribute to any project with contributions from Russia-based developers. In short, I ask if leavers previously had more collaborative ties to developers in other countries than remainers. I find strong evidence that this is the case. In the global collaboration network, 43.0\% of leavers had ties with developers in other countries compared with only 24.3\% of remainers. This presents another way to describe the impact of emigration on the Russian software industry: 11\% of leavers were responsible for 1 in 5 collaborations with to other countries.

I can also test if leavers are more likely to relocate to countries in which they have a previous collaborative tie. 11.5\% of leavers had such a connection. To better understand this quantity and its statistical significance, I compare it against realizations of a null model randomizing leaver destinations. Specifically, I shuffle the set of destination countries across leavers. This randomization preserves the number of developers going to each country. Across 1,000 randomizations, I find that on average only 0.4\% of leavers go to countries with which they previously had a collaborative tie (standard deviation 0.1\%, Z-score vs. observed value $\approx$ 87.6). The observed likelihood is a factor 27 times larger than the null model. I also note that this is likely an underestimate of the true value as the previous dataset could only locate about half of all highly active developers.

\subsection*{Destinations}
Having seen that Russian developers are indeed emigrating, and that leavers are relatively more active and collaborative, I now turn to the question of where these people are going. I compare the top 20 destination countries of geolocatable developers leaving Russia, Belarus, and Ukraine in Table \ref{tab:destinations}. 

Where do developers go? The Russian diaspora is highly dispersed - among confirmed leavers, no single destination receives more than 13\% of developers. The top destinations are a mix of large advanced economies like the US and Germany and nearby countries like Georgia and Armenia. Belarusian leavers, on the other hand, have a much smaller geographic range: over half (53\%) move to the neighboring countries of Poland, Lithuania, Ukraine, or Latvia. Factors like ease of immigration, previous social connections, and economic opportunity at the destination are likely to play key roles in a person's choice of destination. It is unclear how stable these distributions will be as the conflict continues.

\begin{table}
\ra{1.1}  
\begin{tabular}{lrl|lrl|lrl}
\toprule
        \multicolumn{3}{c}{Russia} &  \multicolumn{3}{c}{Belarus}&
         \multicolumn{3}{c}{Ukraine}\\
\midrule
             Destination &  Count & Pct. &             Destination &  Count & Pct. &              Destination &  Count & Pct. \\

\midrule
       United States &    206 &  12\% &         Poland &    179 &  46\% &  United States &     62 &  21\% \\
             Germany &    155 &   9\% &        Georgia &     42 &  11\% &        Germany &     35 &  12\% \\
             Georgia &    127 &   7\% &      Lithuania &     30 &   8\% &         Poland &     32 &  11\% \\
         Netherlands &    108 &   6\% &  United States &     22 &   6\% &         Canada &     27 &   9\% \\
             Armenia &     96 &   6\% &          Spain &     10 &   3\% &         Russia &     20 &   7\% \\
              Cyprus &     89 &   5\% &        Estonia &     10 &   3\% & United Kingdom &     14 &   5\% \\
             Türkiye &     88 &   5\% &        Germany &     10 &   3\% &    Netherlands &     12 &   4\% \\
              Serbia &     78 &   5\% &         Russia &      8 &   2\% &        Czechia &     11 &   4\% \\
      United Kingdom &     65 &   4\% & United Kingdom &      7 &   2\% &          Spain &      7 &   2\% \\
U.A.E. &     65 &   4\% &        Finland &      7 &   2\% &       Slovakia &      6 &   2\% \\
          Kazakhstan &     49 &   3\% &         France &      7 &   2\% &       Portugal &      5 &   2\% \\
             Finland &     38 &   2\% &        Ukraine &      6 &   2\% &        Estonia &      5 &   2\% \\
              Poland &     37 &   2\% &         Cyprus &      5 &   1\% &         Sweden &      5 &   2\% \\
          Montenegro &     35 &   2\% &         Norway &      5 &   1\% &          Italy &      4 &   1\% \\
              Israel &     32 &   2\% &    Netherlands &      5 &   1\% &         France &      3 &   1\% \\
              Canada &     26 &   2\% &       Portugal &      4 &   1\% &        Austria &      3 &   1\% \\
         Switzerland &     26 &   2\% &        Czechia &      4 &   1\% &        Ireland &      3 &   1\% \\
           Indonesia &     25 &   1\% &         Sweden &      3 &   1\% &    Switzerland &      3 &   1\% \\
             Czechia &     24 &   1\% &        Ireland &      3 &   1\% &         Latvia &      2 &   1\% \\
             Estonia &     24 &   1\% &        Türkiye &      3 &   1\% &        Georgia &      2 &   1\% \\
\bottomrule
\end{tabular}
\caption{The top 20 national destinations of geolocatable developers having left Russia, Belarus, and Ukraine between January 2021 and November 2022. Russian developers disperse widely, while a majority of Belarusian developers remain in nearby countries.}
\label{tab:destinations}
\end{table}

For some receiving countries, the influx of developers represents a truly significant shock. In Table 3 I report the receiving countries for which the arrivals represent at least a 5\% increase in the previously observed population of GitHub developers in 2021\footnote{See the count of developers located via GitHub location here: \url{https://github.com/johanneswachs/OSS_Geography_Data/blob/main/data/world_countries_2021.csv}} \cite{wachs2022geography}. I visualize the percent increase in Figure 6. For example, my analysis estimates that the number of active GitHub developers in Georgia has roughly doubled since the Russian invasion of Ukraine.

\begin{table}[]
\ra{1.1}  
\begin{tabular}{lrrr}
\toprule
            Country &  Developers 2021 &  Gain & Pct. Increase \\
\midrule
             Georgia &       180 &   169 &      94\% \\
          Montenegro &        48 &    35 &      73\% \\
              Cyprus &       157 &    94 &      60\% \\
             Armenia &       233 &    97 &      42\% \\
            U.A.E. &       465 &    66 &      14\% \\
          Kazakhstan &       393 &    50 &      13\% \\
              Serbia &       953 &    79 &       8\% \\
          Uzbekistan &       185 &    15 &       8\% \\
          Kyrgyzstan &       125 &     9 &       7\% \\
           Lithuania &       683 &    44 &       6\% \\
             Estonia &       600 &    34 &       6\% \\
\bottomrule
\end{tabular}
  \caption{The receiving countries of Russian and Belarusian developers for which the incoming flow of developers exceeds 5\% of their estimated active GitHub developer population in 2021.}
    \label{tab:top_relative_receivers}
\end{table}

\begin{figure}[t]
    \centering
    \includegraphics[width=0.6\textwidth]{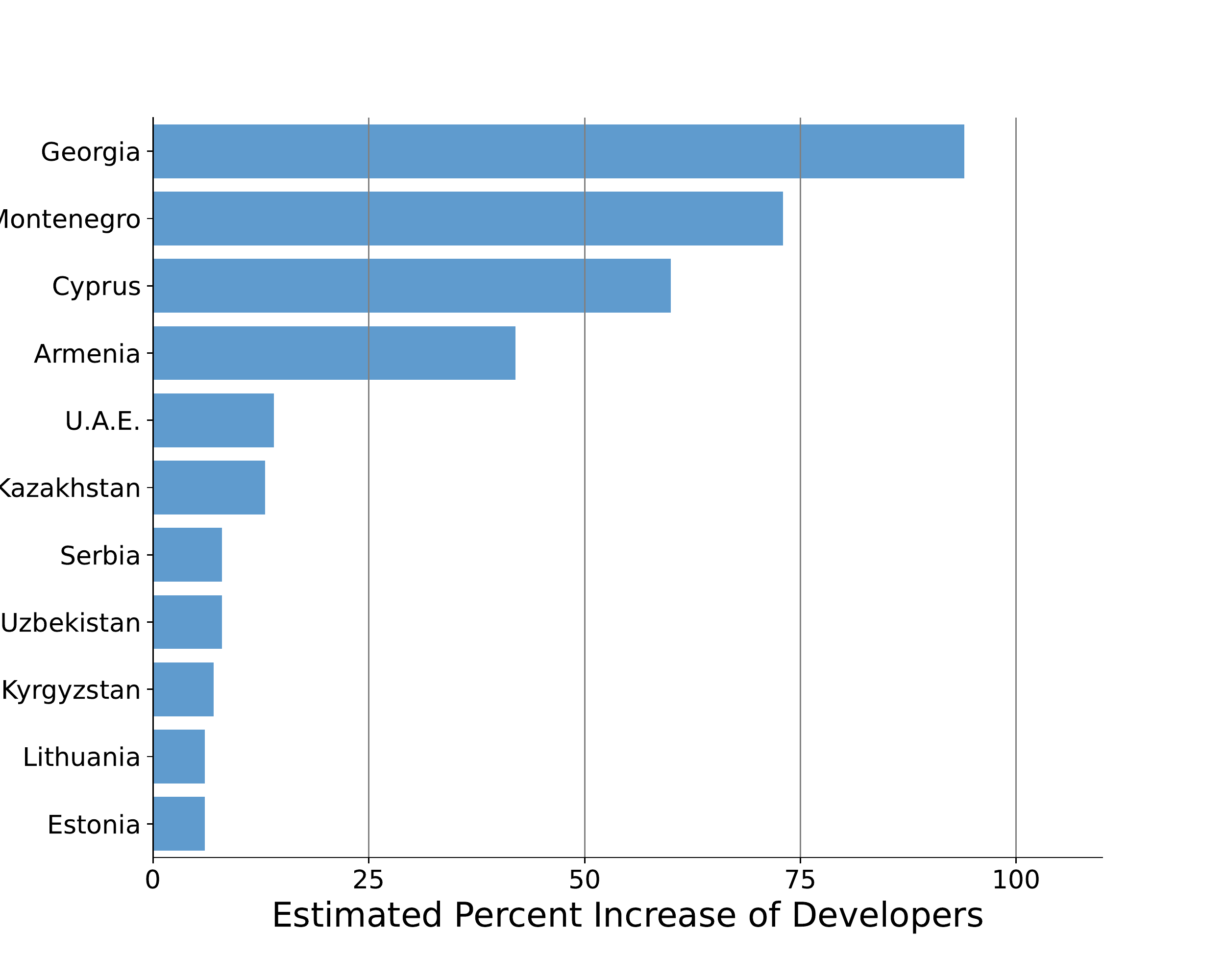}
    \caption{The receiving countries of Russian and Belarusian developers for which the incoming flow of developers exceeds 5\% of their estimated active GitHub developer population in 2021.}
    \label{fig:receiver_bar}
\end{figure}

\section*{Discussion}
In this paper I have shown how digital trace data from GitHub can be used to estimate high skilled emigration during a crisis. The estimates suggest that somewhere between 11-30\% of active Russian and Belarussian OSS developers have left their countries between February 2021 and November 2022. This turnover is more than three times the rate observed in comparable countries from the region not directly involved in the conflict. Those who left Russia are on average much more active OSS developers, and occupy a more central position in the collaboration network. In particular, the 11.1\% confirmed leavers account for 14\% of total Russian code contributions identified in 2019 and 2020. They are also responsible for 20\% of Russia's pre-invasion collaborative ties to other countries. 

To sum up: the invasion of Ukraine has likely been a major push factor, leading to a significant exodus of highly skilled individuals. These individuals with ``upper-tail'' human capital are likely to have significant impact on receiving economies \cite{squicciarini2015human}. For countries like Georgia Montenegro, Cyprus, and Armenia the estimated inflow of developers from Russia and Belarus exceeds 50\% of their previous stock. More generally, these findings suggest the potential of digital trace data to map brain drain flows in crisis situations. Before discussing potential consequences, I now review potential limitations of this work.

\paragraph{Limitations and threats to validity}

As I compare three snapshots of data, there are many ways in which the data may be biased or flawed. For instance I cannot be sure the if the first large migration I observe happened after the invasion or between February 2021 and its start. The comparison with other Eastern European countries mitigates this concern to some extent, as one would expect the evolution of labor mobility to be roughly similar around the region. On the other hand, it may be the case that mobility in the comparison set of countries fell after the invasion, as a conservative response to the war. While countries in the comparison set have diverse exposure to the war, I cannot exclude this possibility as I do not have an earlier estimate of the rate of change of skilled emigration. 

While such a baseline would be valuable, data on historical locations given by developers is not available on GitHub. Indeed, the general absence of data on high skilled worker mobility is one of the motivations of this work. That I observe similar trends in the November snapshot also increases confidence of the validity of the results. I also observe limited migration to Russia: out of more than 28,000 developers in the data excluding Russia, there are only 28 re-locations into Russia by November 2022.

My analysis also takes developer provided locations at face value. GitHub locations are not verified and developers can provide a false location. Some Russia-based developers likely have significant incentive to provide a location outside of Russia for social and economic reasons. For instance, a freelance developer based in ``Moscow'' may have trouble getting clients who are wary of doing business with Russia because of sanctions or social factors. There are also probably fundamental differences between developers who report locations and those who don't. More active developers are probably more likely to disclose more information. This poses limited risk to the analyses as they already focus on quite active contributors (at least 100 commits, reporting similar results in the appendix when filtering at 200 and 500). 

The definition of collaboration used in this work also has limitations. Just because two developers contribute to the same OSS project does not mean they are collaborating in a conventional sense. Nevertheless, the results are robust when considering only projects with at most 10 contributors. Future work may consider collaboration on multiple projects or drill down to the level of co-editing specific files.

Despite these limitations, the weight of the evidence for a significant wave of brain drain merits interpretation of these results in the context of related literature on brain drain. In line with that literature, I now outline potential consequences for the sending and receiving countries.

\paragraph{Consequences}
Russia and Belarus will likely face significant shortages of skilled software developers and experience a long term slow down in technological growth and innovation in the ICT sector. Developers are needed to both create and maintain \cite{lehman1980programs} software systems which are essential to modern economies \cite{holmberg2010maintenance}. Remaining software talent will become more expensive, but it is not clear if higher wages can draw in new developers in the short or medium run, as this kind of work requires lengthy and specialized training. These dampening effects are likely to be long lasting: the effects of human capital loses in World War 2 persisted for decades longer than the effects of damage to physical infrastructure \cite{waldinger2016bombs}. To the extent that the results generalize to other highly-skilled sectors of the Russian economy, the picture of the long run economic development of worsens.

A silver lining of brain drain in general is that leavers tend to stay in touch with their homelands \cite{saxenian2005brain,docquier2012globalization}. They often send remittances, share information about new ideas and opportunities, and act as bridges in collaborations \cite{di2022ties}. In the long run, returning \'emigr\'es can have a significant impact on local economies by bringing new skills and perspectives \cite{hausmann2018welcome}. These virtuous forces only apply if individuals who have left have an interest in remaining connected to their former homes, and perhaps in some day returning. The data presented in this paper contain some hints that many of these emigrations, at least from Russia, may be long lasting. For instance, the United States is by far the most popular destination among developers previously located in Russia, while a majority of developers leaving Belarus remain in the region. Greater distances impose significant costs on collaboration and communication, even in OSS development \cite{takhteyev2010investigating,lima2014coding,fackler2020gravity}. Moreover, the war, ongoing since 2014, has already had a measurable impact on cross-border collaborations in tech \cite{laurentsyeva2019friends}. In short, it is likely that Russia will face many of the negatives of brain drain with few of the positives.

The significant number of software developers arriving to various countries is likely to have an impact on local economies. For example, I estimate that Georgia is receiving as many developers from Russia and Belarus as they previously had in total \cite{wachs2022geography}. If developers settle in regions without significant software presence they could accelerate technical adoption \cite{miguelez2022migrant}. When such industries already exist, they can decisively influence their future development \cite{moser2014german,ganguli2015immigration,juhasz2021explaining}. Places that can attract, retain and effectively integrate these talented individuals will reap substantial dividends. The long run outcomes of this new wave of Russian \'emigr\'es and their influence on their new homes merits continued study.

\section*{Competing interests}
The authors declare that they have no competing interests.

\section*{Availability of data and materials}
An anonymized dataset including only post-processed geolocations is available here: \url{https://github.com/johanneswachs/ru_braindrain_data}. I report robustness tests in a supplementary appendix.

\section*{Author's contributions}
JW is responsible for all content.

\section*{Acknowledgements}
I thank S\'andor Juh\'asz, Zs\'ofia Cz\'em\'an, Gerg\H o T\'oth, B\'alint Dar\'oczy, 
Sergey Alekseev, Frank Neffke, and William Schueller for advice and feedback on preliminary versions of this work.

% if your bibliography is in bibtex format, use those commands:

\bibliography{bmc_article}      % Bibliography file 
\end{document}